\documentclass[aps,prl,showpacs]{revtex4}

\usepackage[dvips]{graphicx}
\usepackage{psfig,epsfig}
\usepackage{amsmath}
\usepackage{amsfonts}
\usepackage{amssymb}
\usepackage{wasysym}

\expandafter\ifx\csname package@font\endcsname\relax\else
 \expandafter\expandafter
 \expandafter\usepackage
 \expandafter\expandafter
  \expandafter{\csname package@font\endcsname}%
\fi

\begin{document}
\def\be{\begin{equation}}
\def\ee{\end{equation}}

\title{Magnitude clustering and  dynamical scaling in 
  trigger models for earthquake forecasting}

\author{Eugenio Lippiello,$^1$ Cataldo Godano$^2$ and Lucilla de Arcangelis$^3$}
\affiliation{$^1$ Physics Department and CNISM, University of Naples
"Federico II", 80125 Napoli, Italy \\
$^2$ Department of Environmental Sciences and CNISM, Second
University of Naples, 81100 Caserta, Italy \\
$^3$ Department of
Information Engineering and CNISM, Second University of
Naples, 81031 Aversa (CE), Italy}

\begin{abstract}
One of the main interests in seismology is the formulation of
models able to describe the clustering in time occurrence of earthquakes.
Analysis of the Southern California Catalog shows magnitude clustering 
in correspondence to temporal clustering. Here we propose a dynamical 
scaling hypothesis in which time is rescaled in terms of
magnitude. This hypothesis is introduced in the context of a
generalized trigger model and 
gives account for clustering in time and magnitude for
earthquake occurrence. The model is able to generate 
a synthetic catalog reproducing magnitude and inter-even time 
distribution of thirty years California seismicity.

\end{abstract}


\maketitle

The great interest in the study of earthquake occurrence is linked to the
challenge of predicting the time, the location and the energy of the
next earthquake. The energy release $E$ in a seismic event can be
expressed by the magnitude $M$ via the logarithm relation
$M\propto\log E$ \cite{kan}, and the magnitude distribution is described by
an exponential law usually referred as the Gutenberg-Richter (GR)
law \cite{gutri}
$P(M)\sim 10^{-bM}$, where $b$ is a parameter close to one.
The logarithm relation leads to a power law behaviour for the
energy distribution, which is generally the signature
of critical phenomena.


It is widely observed that earthquakes tend to occur in bursts.
These bursts start immediately following a large main event,
giving rise to the main-aftershock sequences, described
by the Omori law \cite{Omori}. This states that the number of
aftershocks $n(t)$ decays in time as $n(t)\sim (t+c)^{-p}$
where $p$ is generally close to 1 and  $c$
is an initial time introduced in order to avoid the divergence at
$t = 0$. The most important implication of this law is that we
cannot assume a Poissonian occurrence for earthquakes, namely
characterized by a constant rate of occurrence, but rather a
clustered one.

Another signature of non-Poissonian behaviour for earthquake
occurrence is the complex distribution of the inter-occurrence
times between two successive events. In fact, for a Poissonian
process, this distribution would be an exponential whereas
experimental data exhibit a more complex behaviour \cite{Corral}.
Moreover, one can compute the intertime distribution $D(\Delta
t,M_L)$ where $\Delta t$ is time distance between successive
events occurred inside a finite geographic region and with
magnitude greater than a given threshold $M_L$. Indicating with
$P_C(M)$ the cumulative magnitude distribution inside the
considered region, one observes \cite{Corral,Bak} 
\be
D(\Delta t,M_L)=P_C(M_L) f(P_C(M_L)\Delta t) \label{cor}
\end{equation}
where $f$ is a universal function, independent on $M_L$ and on  
the geographical region. 
The observed universality is a further
signature of criticality and indicates that $D(\Delta t,M_L)$ is an
appropriate quantity to characterize the temporal clustering of
earthquakes.

A widely used approach to earthquakes clustering is provided by
"trigger models" \cite{Vere-Jones}. These assume a Poissonian
occurrence of triggering events, whereas the occurrence of the
"triggered" earthquakes is described in terms of a correlation
function with previous events. Among the trigger
models the Epidemic Type Aftershocks Sequence (ETAS), introduced
by Kagan-Knopoff \cite{Kagan-Knopoff} and developed by
Ogata \cite{Ogata}, describes mainshocks and aftershocks 
on the same footing. More precisely, each earthquake
can generate "its own aftershocks" and furthermore the number of
these aftershocks depends exponentially on the magnitude of the
"main". The model has been
deeply investigated analytically and numerically \cite{Helm}.

In this paper we are interested in the description of temporal
evolution of seismic activity. For this reason we neglect spatial
dependencies and treat seismicity as a stochastic process
$M_i(t_i)$, where $M_i$ is the magnitude of the $i-th$ earthquake
occurred at time $t_i$ inside a large but finite geographic
region. The process is defined by the conditional probability density
$p(M(t) \vert \{M_i(t_i)\})$ to have an earthquake of magnitude $M$
at time $t$ given the history of past events $\{M_i(t_i)\}$.
Here we consider a generalized version of the trigger model by Vere-Jones
\be
p(M(t) \vert \{M_i(t_i)\})=\sum _{i:t_i<t} p(M(t) \vert M_i(t_i))+\mu P(M)
    \label{VJ}
    \ee
where $p(M(t) \vert M_i(t_i))$ is the "two-point" conditional
probability density, $\mu$ is a Poissonian rate and the magnitude
distribution $P(M) \sim 10^{-bM}$ obeys the GR law.
Different forms of $p(M(t) \vert M_i(t_i))$ 
correspond to different models for seismicity.
In the ETAS model one assumes \cite{Ogata}
\be
    p(M(t) \vert M_i(t_i))=P(M)g(t-t_i;M_i)
    \label{FAC}
    \ee
where the propagator $g(t-t_i;M_i)
\propto 10^{\alpha M_i}(t-t_i+c)^{-p}$.
In order to have a normalized probability one must impose
$p>1$.
Moreover, if $\alpha \ge b$ the model presents finite time singularity
unless one assumes a large magnitude cut-off \cite{Kagan}. Alternatively,
one must take $\alpha < b$ as supported by some
experimental observations \cite{Helm2}.

A strong assumption of the ETAS model is the factorization in  Eq.(\ref{FAC}), 
which states that the magnitude of an earthquake
is completely independent on the magnitudes and times of
occurrence of previous events. 
In order to test this assumption with real
seismic data, we observe that the quantity
$P_C(M_L)\sim 10^{-bM_L}$ takes the role of a characteristic time
scale in Eq.(\ref{cor}). Hence, if one considers a subset of $N$
events, the quantity $Q=N/[\sum_i 10^{-bM_i}]$
can be related to the rate of occurrence
$R=N/[\sum_i(t_i-t_{i-1})]$, where the sum is
inside the chosen subset. 
To this extent,
we divide data recorded in the Southern California Catalog 
(1975-2004) \cite {cali} in subsequent sets of $200$ events with 
$M \ge 2.5$ and we
compute the quantities $Q_j$ and $R_j$ inside the j-th subset. If the
magnitude distribution were constant in time, as supposed in Eq. (\ref{FAC}),
$Q_j$ should fluctuate around an average value. Conversely, the
experimental $Q_j$ displays scattered and narrow peaks (Fig.1a).
Interestingly, these peaks are closely located to peaks in the $R_j$
distribution. It is well known that peaks of $R_j$ are located
soon after main-shocks and indicate the presence of main-after
shock sequences. Fig.1a, then, shows that in subsets of the catalog 
where activity has an higher rate, 
the probability to have large magnitude events is also raised.
This aspect can be directly investigated by computing
the
%
cumulative magnitude distribution $P_C(M)$ only inside the ensemble of 
main-aftershock sequences.
Considering only sequences with main-shock magnitude  $M \ge 6$, 
one obtains that $P_C(M)$ exhibits a GR behaviour with a best fit $b$-value
$b=0.75$, lower than the $b$-value obtained for the whole catalog
($b=0.95$) within 
the $95\%$ significativity level.
This result further supports the idea that large earthquakes not only
produce the clustering in time described by the Omori law, but
also a clustering in magnitude. The ETAS model does not take into
account this last physical mechanism.
\begin{figure}
\includegraphics[width=8cm]{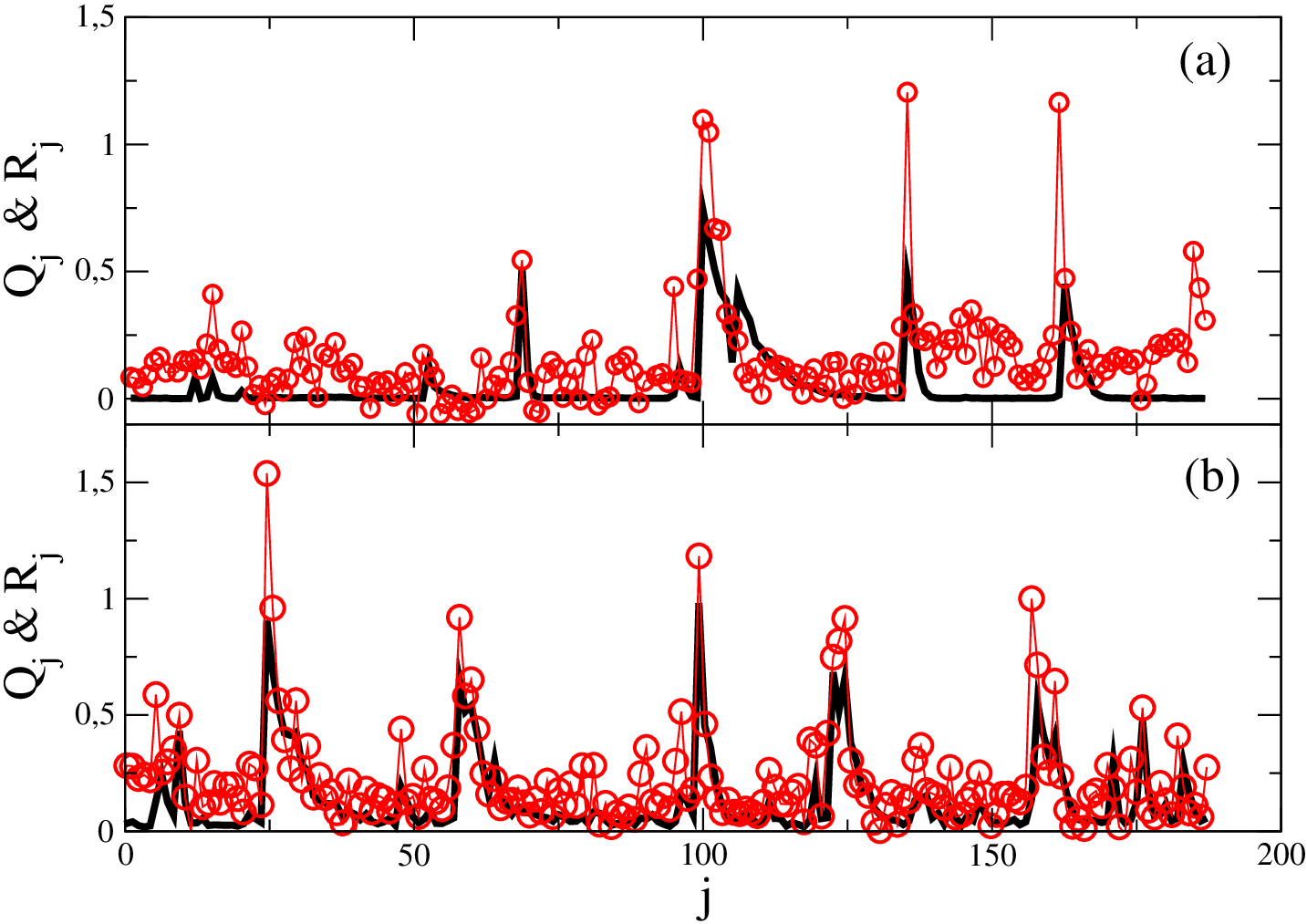}
\caption{(Color online) (a) Experimental distributions of the quantities $Q_j$
  (red $\circ$) and $R_j$ (black continuous line) for the California Catalog.
$R_j$ is measured in unit of $(6 hours)^{-1}$ and in order to 
improve the comparison $Q_j$ is vertically shifted by the constant
  amount $-1.5$.
 Peaks for
  $R_j$ indicate main-after shock sequences. (b) Numerical
  distributions of $Q_j$ and $R_j$ from Eq.(\ref{fz2}). 
}\end{figure}

In order to include the magnitude clustering within a trigger
model approach, we propose a dynamical scaling hypothesis: the magnitude
difference $M_i-M_j$ fixes a characteristic time scale
$ \tau_{ij}=k 10^{\frac{b}{p}(M_j-M_i)}$
so that the conditional
probability is magnitude independent when times are rescaled by
$\tau_{ij}$ and $k$ is a constant measured in seconds
\be
    p(M_i(t_i) \vert M_j(t_j)) =F \left [
    \frac{(t_i-t_j)}{\tau_{ij}}\right ]
    \label{scaling}
    \ee

Let us then
consider the probability to have an event of magnitude $M$ at time
$t$ given a triggering event at time $t_0$ of arbitrary
magnitude $M_0$,
    $p(M,t-t_0)=\int dM_0 P(M,t \vert M_0,t_0) P(M_0)$.
Assuming the GR law for  $P(M_0)$ and using Eq.(\ref{scaling}), one
finds
\be
    p(M,t-t_0)=\frac{10^{-b M}}{(t-t_0)^p} \int_0^{10^{bM}(t-t_0)^p}
    F(z^{1/p})dz.
    \label{Pm1}
    \ee
From this equation we obtain both the GR and Omori law
independently of the specific form of $F(z)$ provided that the
appropriate constraints are imposed at small and large $z$. In
fact, assuming that the conditional probability (6) is maximal
soon after the triggering event, must be $F(0)>0$.
Furthermore, in order to have normalized distributions, the
conditional probability must decay to zero for large time
separation and a constraint on the behaviour of $F(z^{1/p})$ must be
imposed at large $z$, namely a decay faster than $1/z$. Because of
this constraints, the integral in the rhs of Eq.(\ref{Pm1}) is a
constant for large $t-t_0$, and the GR and Omori law directly
follows from Eq.(\ref{scaling}).
The above observation suggests that statistical features of the
trigger model can be independent on the detailed form of $F(z)$
once the scaling Eq.(\ref{scaling}) is assumed. This hypothesis
together with the relationship between numerical and experimental
behaviour can be directly tested  in numerical simulations.

In a numerical protocol one assumes at initial time $t_0=0$ a
single event of arbitrary magnitude chosen in a fixed range
$[M_{min},M_{max} ]$. Time is then increased of a unit step
$t=t_0+1$, a trial magnitude is randomly chosen in the interval
$[M_{min},M_{max} ]$ and Eq.(\ref{VJ}) gives the probability to
have an earthquake in the time window $(t_0,t_0+1)$. If this
probability is larger than a random number between $0$ and $1$, an
earthquake takes place, its magnitude and time of occurrence are
stored and successively used for the evaluation of probability for
future events. Time is then increased and in this way one
constructs a synthetic catalog of $N_e$ events. The term $\mu$ in
Eq.(\ref{VJ}) represents an additional source of earthquakes
Poissonian distributed in time with a magnitude chosen from the GR
distribution with $b=0.8$.

Following this protocol, we generate sequences of $15000$ events
using a power law form for $F(z)$
\be
   F(z)=\frac{A}{z^\lambda+\gamma}
\label{fz}
\ee
and then
we compute the numerical distributions $D(\Delta t,M_L)$
and $P(M)$. These distributions are compared
with the experimental data from the Southern California Catalog.
For different values of $\lambda$, it is always possible to find a set of
parameters ${A,\gamma,b/p,\mu}$ such that numerical data reproduce, on average,
the statistical features of earthquake occurrence both in time and in
magnitude. The parameter $k$ is fixed a posteriori in order to obtain
the collapse between numerical and experimental data.

In Fig.2 we plot the experimental and numerical $D(\Delta t,M_L)$
considering two different values of $\lambda$
$(\lambda=1.2$ and $5)$ and $M_L$ ($M_L=1.5$ and $2.5$).
In the inset we also present the magnitude distributions.
Data for different values of the parameters follow a universal curve and the
same collapse is obtained for other values of $\lambda>1$.
The accordance between experimental and numerical curves indicates that
the hypothesis of dynamical scaling is able to reproduce
two fundamental properties of seismic occurrence, namely the GR law
and Eq. (\ref{cor}), independently of the details of $F(z)$ \cite{nota}.

The ETAS model is a particular case of Eq.(\ref{fz}) corresponding
to $\gamma=0$ and $\lambda \simeq 1$. We want to stress the important
difference due to the presence of a non-zero $\gamma$. From a
mathematical point of view, the constant $\gamma$ avoids the finite
time singularity of the ETAS model with $\alpha=b$ discussed
previously \cite{Helm}. From a physical point of view, the constant $\gamma$ gives
rise to the observed clustering in magnitude. Indeed, for a given
mainshock of magnitude $M_j$ at time $t_j$, at each time
$(t_i>t_j)$ it is possible to define a sufficiently large
magnitude difference
$\Delta M$ such that, if $M_j - M_i > \Delta M$, we have
that $z^\lambda$ is negligible with respect to $\gamma$ and therefore
$F[(t_i-t_j)/\tau_{ij}]\simeq A/\gamma$. In other words after a large
event, small earthquakes tend to be equiprobable.

\begin{figure}
\includegraphics[width=8cm]{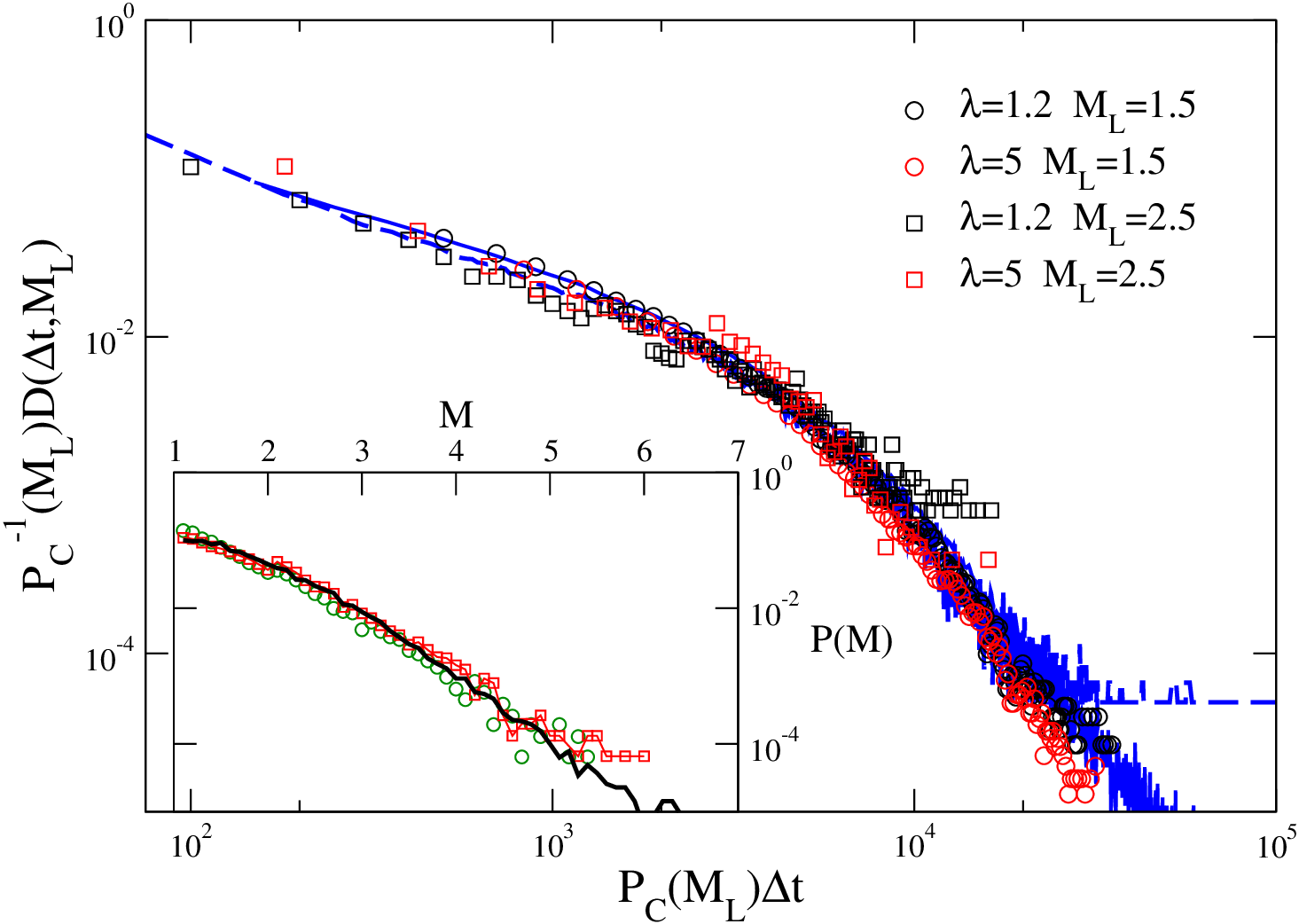}
\caption{(Color online) 
The intertime distribution obtained using Eq. (\ref{fz}),
  with two different values of $\lambda=1.2,5$ and $M_L=1.5,2.5$. 
Continuous and broken curve are the experimental $D(\Delta t,M_L)$
with $M_L=1.5$ and $M_L=2.5$ respectively. 
For $\lambda=1.2$ we set $k=210sec$, $A=1.4 10^{-4}sec^{-1}$, $\mu =4
  10^{-7}$, $\gamma=1$. 
For $\lambda=5$ we set $k=420sec$, $A=1.9 10^{-4}sec^{-1}$, $\mu =1.5
  10^{-6}$, $\gamma=0.1$. 
In the inset the 
magnitude distribution of the experimental catalog (black line) 
and numerical catalog with  $\lambda=1.2$ (red $\circ$) and
  $\lambda=5$ (green $\square$). 
}
\end{figure}

\begin{figure}
\includegraphics[width=8cm]{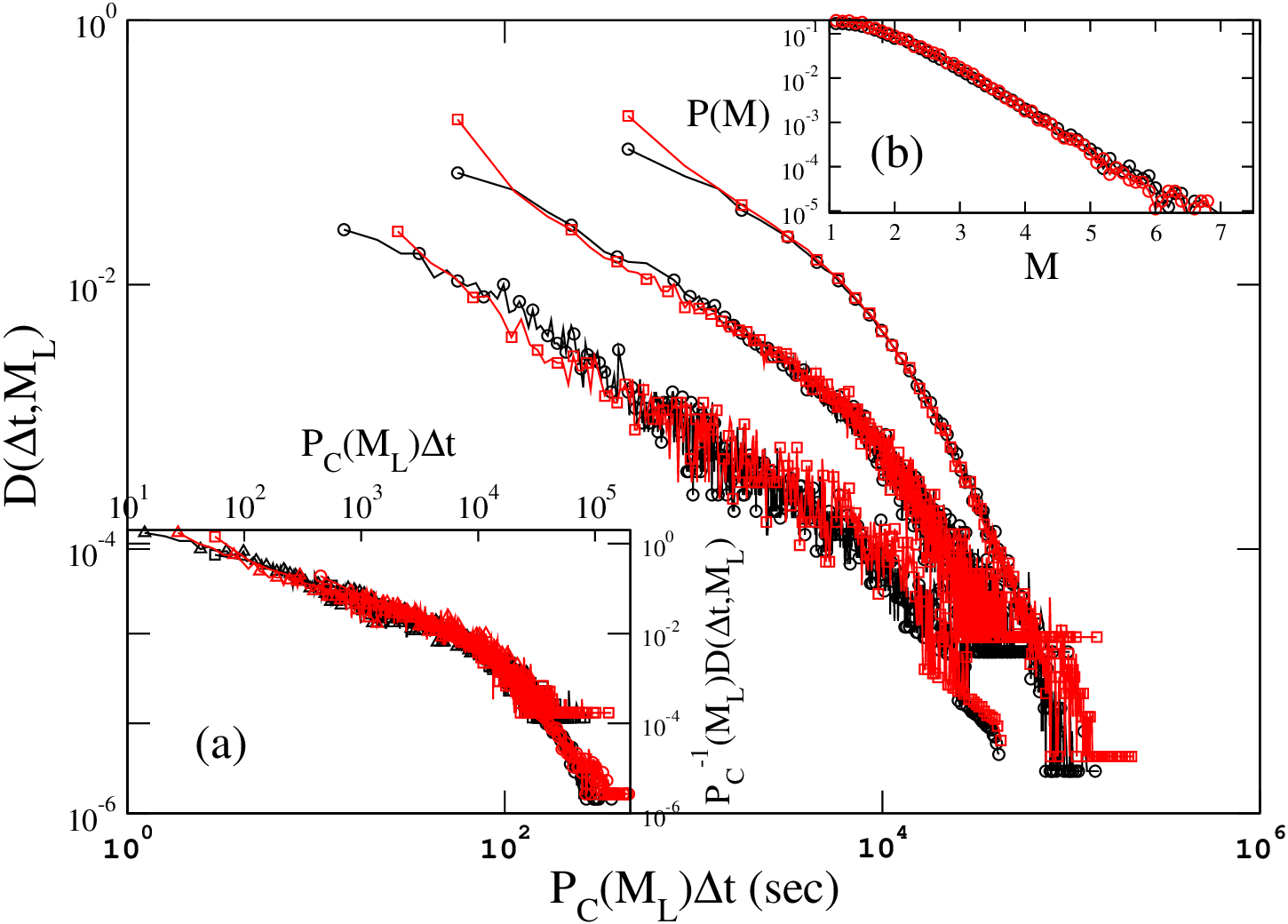}
\caption{(Color online) The intertime distribution as a function 
of $\Delta t P_c(M_L)$
obtained using Eq.(\ref{fz2}) (black circle {\Large $\circ$}) and
compared with the experimental distributions (red $\square$) 
for three different
values of $M_L$ ($M_L=1.5,2.5,3.5$, from top to bottom). 
We set $k=4.9*10^4 sec$, $A=6.1*10^{-5}sec^{-1}$, $\mu =2*10^{-5}$, 
$\gamma=0.1$. 
In inset (a) the scaling 
behaviour as in Eq.(\ref{cor}) and in inset (b) the experimental (red) and 
numerical (black) magnitude distribution.
} \label{fig3}
\end{figure}

We have also performed more extensive simulations using a
different expression for $F(z)$
\be
    F(z)=\frac{A}{e^{z}-1+\gamma}
    \label{fz2}
    \ee
Eq.(\ref{fz2})
states that two events of magnitude $M_i$ and $M_j$ are correlated
over a characteristic time $\tau_{ij}$ and become independent when
$t_i-t_j
>\tau_{ij}$. As a consequence only a small fraction of previous events
can affect the probability of future earthquakes so that, after a
certain time, Earth crust loses memory of previous seismicity.
This aspect is perhaps more realistic with respect to the idea,
contained in a power law correlation, that  events are all
correlated with each other and also gives rise to important implications
for seismic forecasting. The construction of seismic catalogs,
indeed, dates back to about 50 years, and according to
Eq.(\ref{fz2}) one can have good estimates of seismic hazard
without considering previous seismicity. This is no longer true if
one assumes a power law time decorrelation of the type (\ref{fz})
especially for small values of $\lambda$. We want also to point
out that a general state-rate formulation \cite{Die} gives rise to
correlations between earthquakes that decay exponentially in
time. We finally observe, that taking into account only a fraction
of previous events in the evaluation of conditional probabilities, the
numerical procedure considerably speeds up.  In the case of long
temporal correlation CPU time grows with the number of events as
$N_e^2$, whereas in the case of an exponential tail the growth is
linear in $N_e$. For this reason, assuming the functional form
(\ref{fz2}) one can simulate very large sequences of events. In
particular for a different choice of parameters, one can
construct synthetic catalogs containing the same number of events
($ N_e=245000$ with $M \ge 1.5$) of the experimental
California Catalog. In fig. \ref{fig3} we compare numerical and
experimental distributions $D(\Delta t,M_L)$ for three different
values of $M_L$. For each value of $M_L$, the
numerical curve reproduces the experimental data and obviously
fulfill Eq.(\ref{cor}) (inset (a) in Fig.(\ref{fig3})). Also the
numerical  magnitude distribution $P(M)$ is in very good agreement
with the experimental one (inset (b) in Fig.(\ref{fig3})). 
Finally, evaluation of quantities $Q_j$ and $R_j$ for the synthetic catalog
leads, as expected, to the same clustering behaviour as for
experimental data (Fig.1b).
After fixing $k$, we express numerical time unit in seconds and   
we observe that numerical catalog corresponds to a period of about 
$9.9*10^9 sec \simeq 30$ years. Therefore our model is able to construct a
synthetic catalog covering about $30$ years that contains about
the same number of events and displays the same statistical
organization in  magnitude and time of occurrence as real
California Catalog. The high efficiency of the model in
reproducing past seismicity indicates that the model is a good
tool for earthquake forecasting. In fact, given a seismic history, 
Eq.(\ref{VJ}) together with Eq.s(\ref{scaling}, \ref{fz2})
gives the probability to have an earthquake of magnitude $M$ at
time $t$ inside a considered geographic region. Our approach is different
from the Reasenberg-Jones method \cite{RJ}, which is
currently used for evaluation of seismic hazard. This method
is based on the
generalized Omori law that gives for the rate of occurrence of 
magnitude $M$ aftershocks, $P(t,M)=\widehat A 10^{\widehat b (M-M_{M})}
(t-t_{M}+\widehat c)^{-\widehat p}$, where $t_M$ and $M_M$ are the 
time of occurrence and magnitude of the main-shock. The starting set of
parameters ($\widehat b,\widehat p,\widehat c,\widehat A$) is
estimated from previous seismic sequences, and then their value is continuously
updated as soon as new data become available. However, strong fluctuations 
in the magnitude distribution observed in Fig.1 suggest that the extrapolated 
$\widehat b$ from the previous subset may not give the correct value to use for
event forecasting. Furthermore, one has an improving  parameters
estimation as the sequence evolves, but at the same time hazard is
decreasing. Conversely in our model parameters $(A,b/p,\gamma)$
are evaluated on the basis of the entire history of $245,000$ events  
leading to a more precise estimation. Nevertheless, due to the stochastic 
nature of the process, one observes
fluctuations of $\widehat b$ and $\widehat p$
from one sequence to the other (Fig.1b).
Our model, furthermore, also allows hazard estimation outside the
Omori sequence and therefore long term forecasting.

We finally observe that also spatial distributions of seismic
events reveal some kinds of scale invariance \cite{spatial,pacz,god}. These
indicate that also spatial distribution originates from a critical
behaviour of the Earth crust suggesting that a dynamical scaling
hypothesis as in Eq.(\ref{scaling}) can also work if one
appropriately introduces spatial dependencies. In this way it would be possible
to construct seismic hazard maps.

{\small Acknowledgements. This work is part of the project of the
Regional Center of Competence "Analysis and Monitoring of Environmental
Risk" supported by the European Community on Provision 3.16.
This research was also supported by EU Network Number
MRTN-CT-2003-504712, MIUR-PRIN 2004, MIUR-FIRB 2001.}


\begin{thebibliography}{}
\bibitem{nota}
A condition on $\lambda$ should be imposed from the observation that,
taking a single event at $t=t_0$ and neglecting the Poissonian term in
Eq. (\ref{VJ}) one obtains a power law $P(t-t_0) \sim (t-t_0)^{-\lambda p}$.
Hence, one is tempted to  fix $\lambda \simeq 1$ in order to reproduce the
Omori law. Nevertheless considering
$\mu > 0$, also for $\lambda$ different than one, one can recover
Omori law with $p$ belonging to the experimental range.
This observation, for instance, can be indirectly extracted from the
short time behaviour of $D(\Delta t,M_L)$ \cite{Mega}.
\end{thebibliography}
\end{document}